\documentclass[11pt]{article}
\usepackage{moriond,epsfig}
\def\be{\begin{equation}}
\def\ee{\end{equation}}
\def\bea{\begin{eqnarray}}
\def\eea{\end{eqnarray}}

\RequirePackage{xspace}
\def \gev          {\ensuremath{\mathrm{\,Ge\kern -0.1em V}}\xspace}
\def \mev          {\ensuremath{\mathrm{\,Me\kern -0.1em V}}\xspace}
\def \kev          {\ensuremath{\mathrm{\,ke\kern -0.1em V}}\xspace}
\def \ev           {\ensuremath{\mathrm{\,e\kern -0.1em V}}\xspace}
\def \gevc         {\ensuremath{{\mathrm{\,Ge\kern -0.1em V\!/}c}}\xspace}
\def \mevc         {\ensuremath{{\mathrm{\,Me\kern -0.1em V\!/}c}}\xspace}
\def \gevcc        {\ensuremath{{\mathrm{\,Ge\kern -0.1em V\!/}c^2}}\xspace}
\def \mevcc        {\ensuremath{{\mathrm{\,Me\kern -0.1em V\!/}c^2}}\xspace}

\def \Vub {\ensuremath{V_{ub}}\xspace}

\def \Vcb {\ensuremath{V_{cb}}\xspace}

\def \CP{\ensuremath{C\!P}\xspace}

\def\Bbar  {\kern 0.2em\overline{\kern -0.2em B}{}\xspace}

\def\Dbar  {\kern 0.2em\overline{\kern -0.2em D}{}\xspace}

\def\Kbar  {\kern 0.2em\overline{\kern -0.2em K}{}\xspace}

\def \babar {{\it Babar}~}
\def \belle {{\it Belle}~}
\def \utfit {{\bf UT}{\it fit}~}

\begin{document}
\vspace*{4cm}
\title{Status of the Unitarity Triangle analysis in \utfit}

\author{V. SORDINI, on behalf of the \utfit collaboration}

\address{ETH - Zurich}

\maketitle


\abstracts{We present here the status of the Unitarity 
Triangle analysis by the \utfit collaboration in and 
beyond the Standard Model.}

\section{Inputs to Unitarity Triangle Analysis}

Many experimentally accessible quantities are related to the angles and 
sides of the Unitarity Triangle (UT) and their measurement can hence put constraints 
on the UT plane coordinates ($\bar\rho$, $\bar\eta$). 

The main experimental and theoretical inputs to the UT 
Analysis performed by the \utfit collaboration are summarized in tables 
\ref{tab:exp_inputs} and \ref{tab:lattice_inputs}, respectively.
The choice of lattice QCD quantities used in UTfit is motivated in 
\cite{ref:UTfit_lattice}.  
The relation between $\bar\rho$ and $\bar\eta$ and the measured quantities 
is discussed for example in \cite{ref:UTfitBASE}. 

\begin{table*}[h]
\begin{center}
\begin{tabular}{|l||c|c|c|}
\hline
Input & Source & Value & Reference \\
\hline
$|V_{ud}|$ &             Nuclear decays              & $0.97418 \pm 0.00026$                      & \\
$|V_{us}|$ &             SL kaon decays              & $0.2246 \pm 0.0012$                        & \\
$|V_{cb}|$ exclusive [$\times 10^{-3}$]&    SL charmed $B$ decays       & $(39.2 \pm 1.1 )$ & \cite{ref:UTfit_lattice}\\
$|V_{cb}|$ inclusive [$\times 10^{-3}$]&    SL charmed $B$ decays       & $(41.68 \pm 0.39 \pm 0.58$ & \cite{ref_hfag}\\
$|V_{ub}|$ exclusive [$\times 10^{-3}$]&    SL charmless $B$ decays     & $(3.50 \pm 0.4)$  & \cite{ref:UTfit_lattice} \\
$|V_{ub}|$ inclusive [$\times 10^{-3}$]&    SL charmless $B$ decays     & $(3.99 \pm 0.15 \pm 0.40)$  & \cite{ref_hfag}\\
${\cal B}(B^+ \to \tau^+ \nu)$ & Leptonic $B$ decays &$(1.73 \pm 0.35) \times 10^{-4}$ & \cite{ref_btaunu}\\
$\Delta m_s$ &  $B_s \bar{B}_s$ mixing               & ($17.77 \pm 0.12$) ps$^{-1}$   & \cite{ref_hfag}\\
$\Delta m_d$ &  $B_d \bar{B}_d$ mixing               & ($0.507 \pm 0.005$) ps$^{-1}$  & \cite{ref_dms}\\
\hline
$|\epsilon_K|$ [$\times 10^{-3}$]& $K \bar{K}$ mixing                  & $(2.232 \pm 0.007)$         & \\
$\sin 2\beta$ & $B\to J\psi$               &   $0.668 \pm 0.028 \pm 0.012$ & \cite{ref:sin2bTh}\\
${\cal B}$, ${CP}$ parameters \,\,& $B \to \pi \pi,\, \rho \rho,\, \rho \pi$ decays  &- & \cite{ref_hfag} \\
($x^{\pm}, y^{\pm}$), ${\cal B}$, $A$ & $B \to D^{(*)0}K^{(*)\pm}$ (GGSZ, GLW, ADS)  &- & \cite{ref_hfag}\\
\hline
\end{tabular}
\end{center}
\caption{Most relevant inputs for the global UT analysis.}
\label{tab:exp_inputs}
\end{table*} 

\begin{table*}[htb]
\begin{center}
\begin{tabular}{|l||c|c|}
\hline
Input & Value & Reference \\
\hline
$f_{B_s}$                     & $(245 \pm 25)$ MeV     & \cite{ref:UTfit_lattice}\\
$\hat{B}_{B_s}$               & $1.22 \pm 0.12$     & \cite{ref:UTfit_lattice}\\
$f_{B_s}/f_{B_d}$              & $1.21 \pm 0.04$  & \cite{ref:UTfit_lattice}\\
$\hat{B}_{B_s}/\hat{B}_{B_d}$  & $1.00 \pm 0.03$     & \cite{ref:UTfit_lattice}\\
$B_K$                         & $0.75 \pm 0.07$    & \cite{ref:UTfit_lattice}\\
\hline
\end{tabular}
\end{center}
\caption{Phenomenological quantities obtained from the 
Lattice QCD calculation.}
\label{tab:lattice_inputs}
\end{table*}

\section{Results of the Unitarity Triangle Analysis in the Standard Model}

\begin{figure}[h]
\begin{center}
\epsfig{file=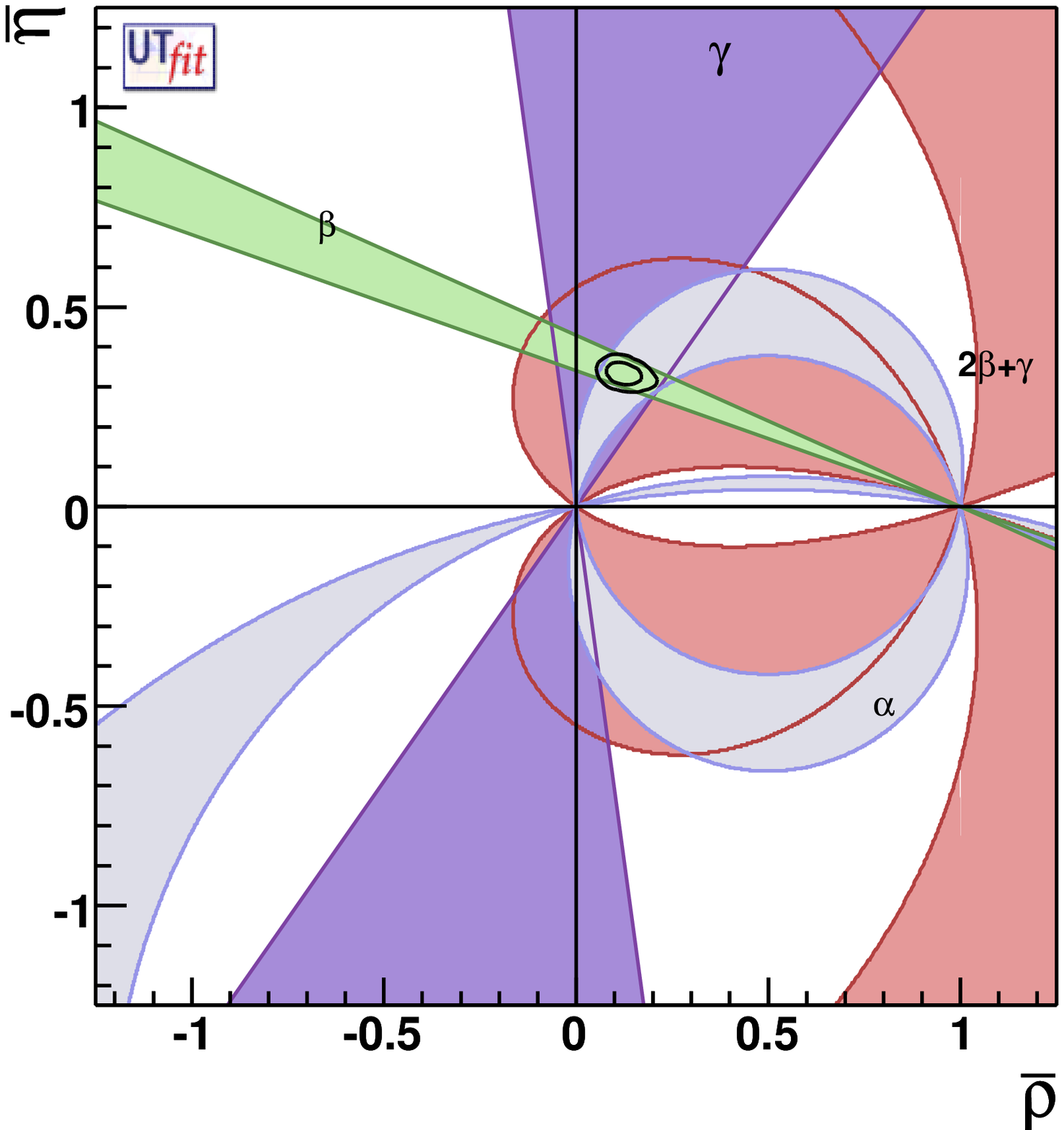,height=5.cm}
\epsfig{file=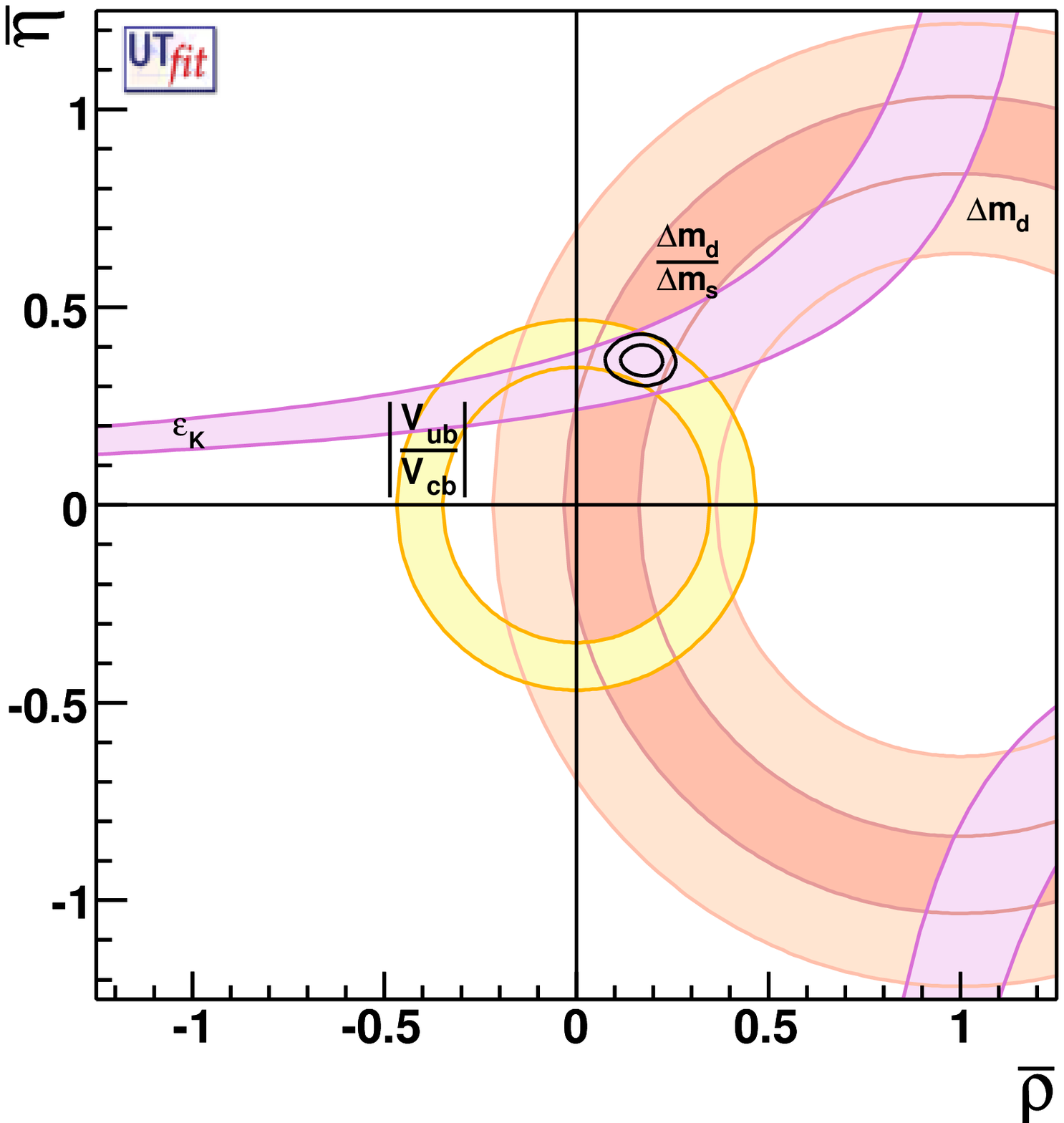,height=5.cm}
\epsfig{file=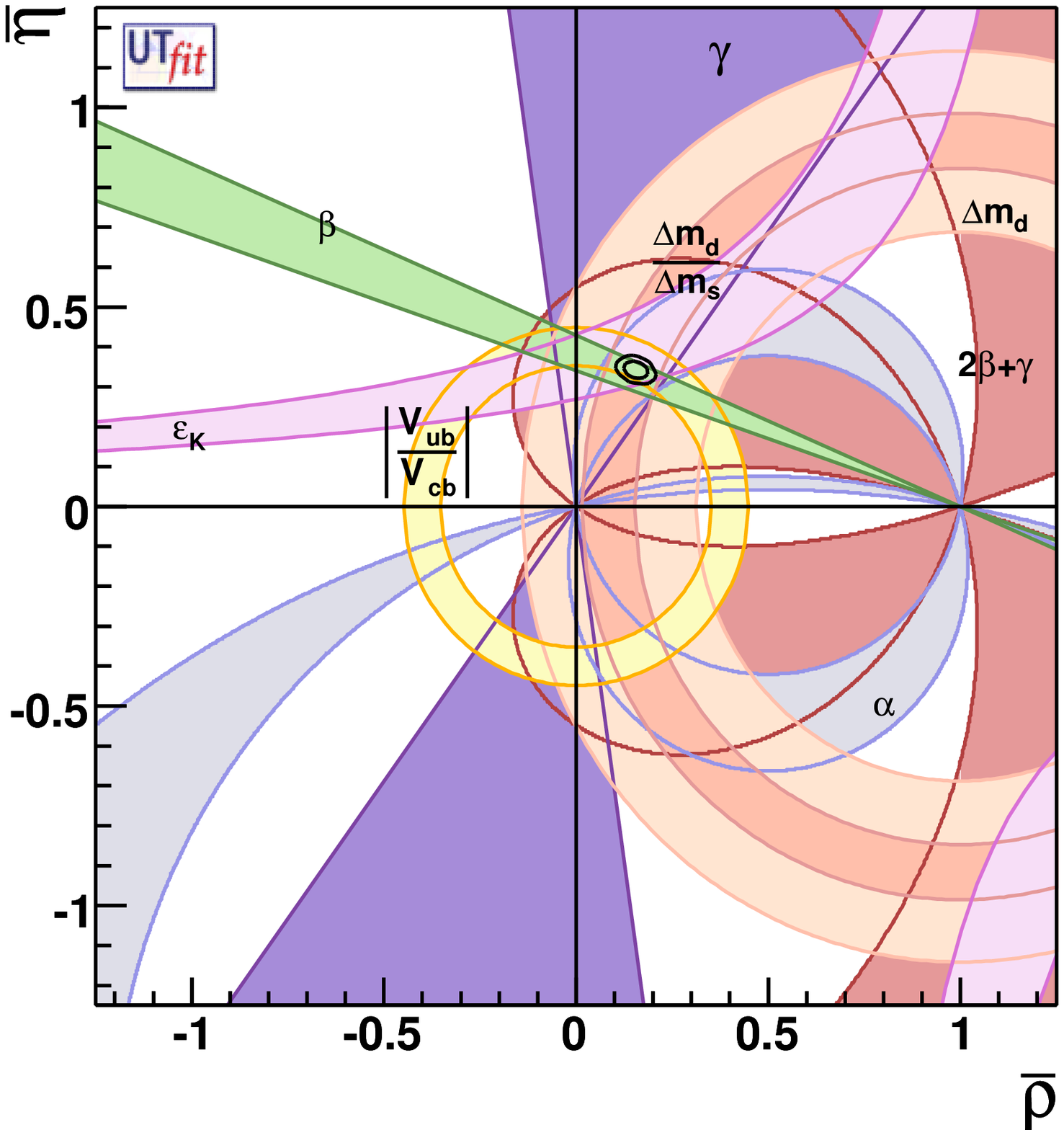,height=5.cm}
\end{center}
\caption{Allowed regions for $\overline{\rho}-\overline{\eta}$, as given by different 
sets of measurements: $|V_{ub}|/|V_{cb}|$, $\Delta m_d$, $\Delta m_s$ and 
$\epsilon_K$ (left plot); $\alpha$, $\sin 2\beta$, $\gamma$, $2\beta+\gamma$, $\beta$ 
and $\cos 2\beta$ (center plot); all these measurements combined (right plot). 
The closed contours show the $68\%$ and $95\%$ probability regions for the triangle apex, while 
the colored zones are the $95\%$ probability regions for each constraint. \label{fig:Ut_angVSoth}}
\end{figure}

The UT analysis performed by \utfit determines the region in which the apex of the 
UT has to be with a given probability \footnote{There we follow a bayesian approach. 
More details are given in \cite{ref:UTfitBASE}}.  
The increasing precision of the measurements and of the theoretical calculations 
in the last twenty years significantly improved the knowledge on the allowed region for 
the apex position $(\overline{\rho},\overline{\eta})$.  
The measurements of \CP-violating quantities from the $B$-factories are nowadays 
so abundant and precise that the CKM parameters can be constrained using only 
the determination of the UT angles, as can be seen in Fig.~\ref{fig:Ut_angVSoth}, 
left plot.
On the other hand, an independent determination can be obtained using experimental 
information on \CP-conserving processes ($\frac{|V_{ub}|}{|V_{cb}|}$ 
from semileptonic $B$ decays, $\Delta m_d$ and $\Delta m_s$ from the $B_{d} - \bar{B}_{d}$ and 
$B_{s} - \bar{B}_{s}$ oscillations) and the direct \CP violation measurements in 
the kaon sector, $\epsilon_K$ (see Fig.~\ref{fig:Ut_angVSoth}, 
center plot). This was indeed the strategy used to predict the value 
of $\sin2\beta$ before the precise \babar and \belle measurements \cite{s2bPrediction}.  
In Fig. \ref{fig:Ut_angVSoth}, right plot, we show the allowed regions for $\bar{\rho}$ and 
$\bar{\eta}$, as given by all the available measurements.  

\begin{table*}[h]
\begin{center}
\begin{tabular}{|l||c|c|c|}
\hline
Parameter & Angles measurements  & $V_{ub}/V_{cb}$, $\Delta m_d$, $\Delta m_s$, $\epsilon_K$ & All \\
\hline
$\bar\rho$ & $0.120 \pm 0.034$ $[0.053,0.194]$ & $0.175 \pm 0.027$ $[0.119,0.228]$& $0.154 \pm 0.022$ $[0.110,0.198]$\\
$\bar\eta$ & $0.334 \pm 0.020$ $[0.296,0.375]$& $0.360 \pm 0.023$ $[0.316,0.406]$& $0.342 \pm 0.014$ $[0.315,0.371]$\\
\hline
\end{tabular}
\end{center}
\caption{Values obtained at 68\%[95\%] probability for $\bar\rho$ and $\bar\eta$ 
from a UT analysis using only angles measurement (first column, labeled "angles") are 
compared with the ones obtained from a UT analysis using semileptonic $B$ 
decays measurements, $\Delta m_d$, $\Delta m_s$ and $\epsilon_K$ 
(second column, labeled "others"). In the third column are also shown the 
results from the complete UT analysis, using all the available measurements.}
\label{tab:rhoeta_all}
\end{table*} 

It can be observed that the Standard Model (SM) description of \CP violation through the 
CKM matrix appears very successful and able to account for all the measured observables 
up to the current precision. 
In this situation, any effect from physics beyond the SM should appear 
as a correction to the CKM picture. 
These remarks do not apply to the observables that have no or very small impact on 
$\bar\rho$ and $\bar\eta$, as the $B_s$ mixing phase, which will be discussed in 
section \ref{sec:NP}. 

\section{Compatibility within different measurements}
\label{sec:compatibility}

We quantify the agreement among all the measured quantities is quantified 
using the {\it compatibility plots} \cite{Bona:2005vz}.  
The indirect determination of a particular quantity is obtained performing the full 
UT, including all the available constraints except from the 
direct measurement of the parameter of interest.  
This fit gives a prediction of the quantity, assuming the validity of the SM. 
The comparison between this prediction and a direct measurement can thus quantify 
the agreement of the single measurement with the overall fit and possibly reveal 
the presence physics phenomena beyond the SM.  
Given the present experimental measurements, no significant deviation from the 
CKM picture is observed.  

\begin{figure}[h]
\begin{center}
\epsfig{file=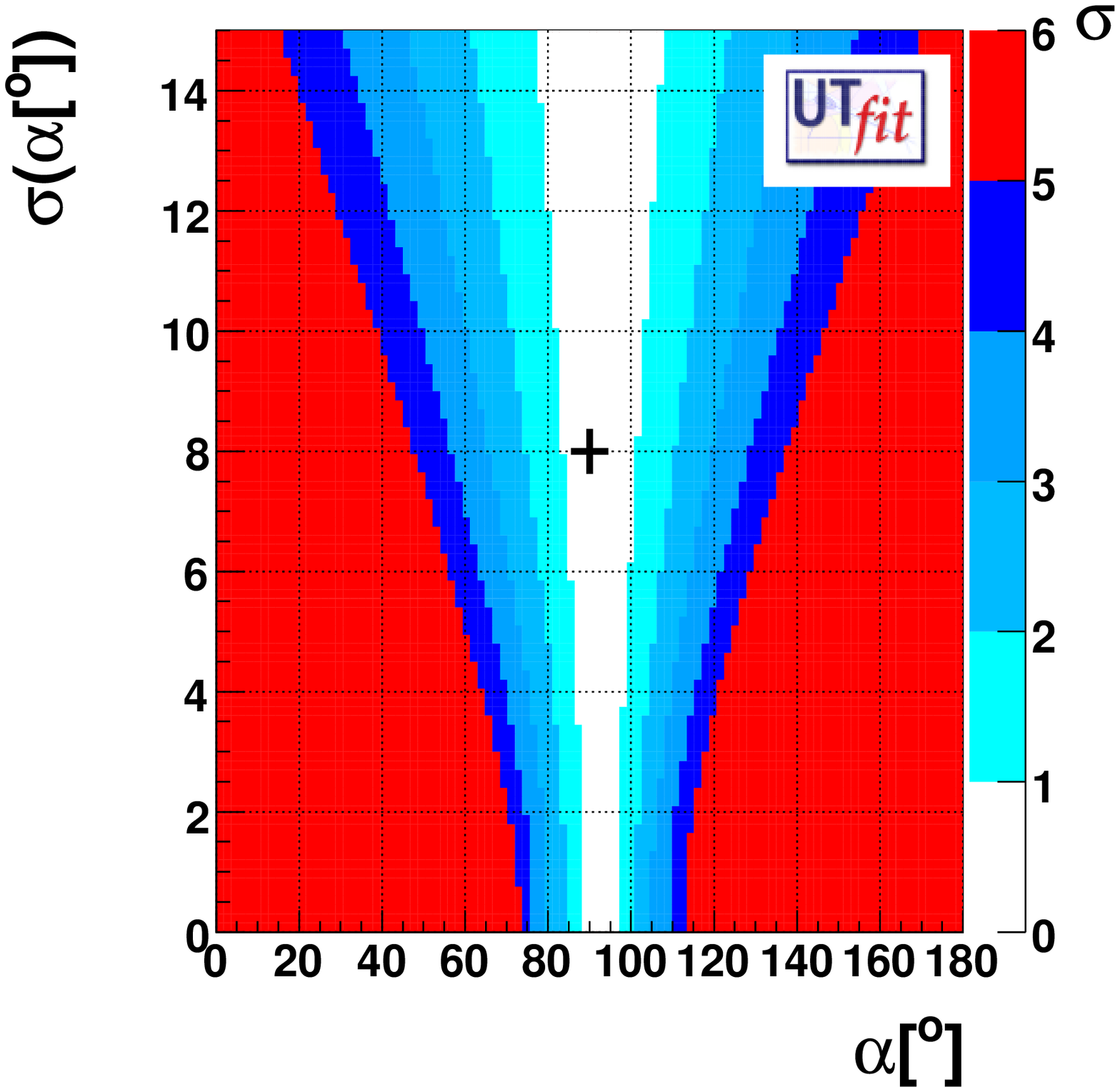,height=3.5cm}
\epsfig{file=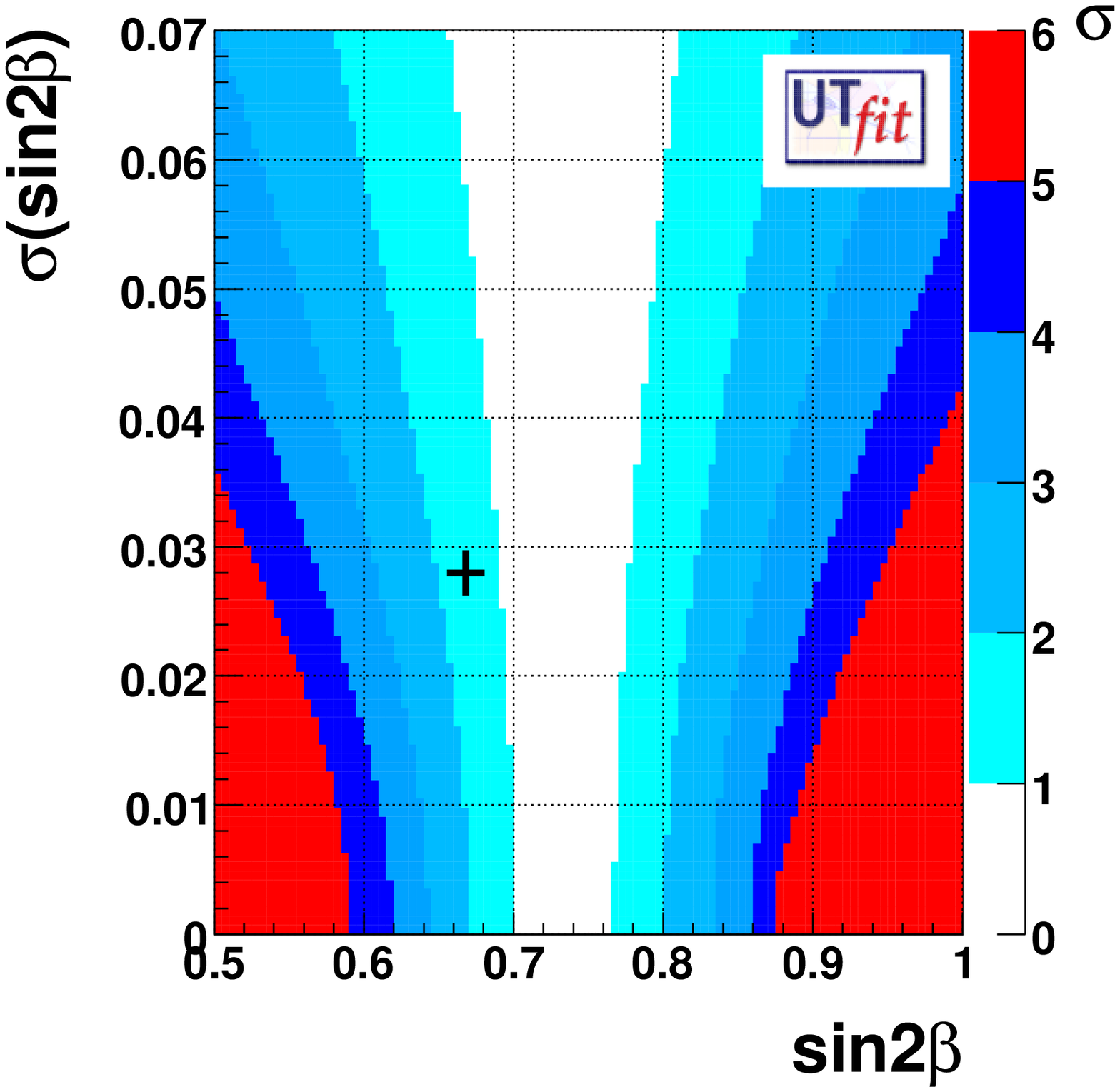,height=3.5cm}
\epsfig{file=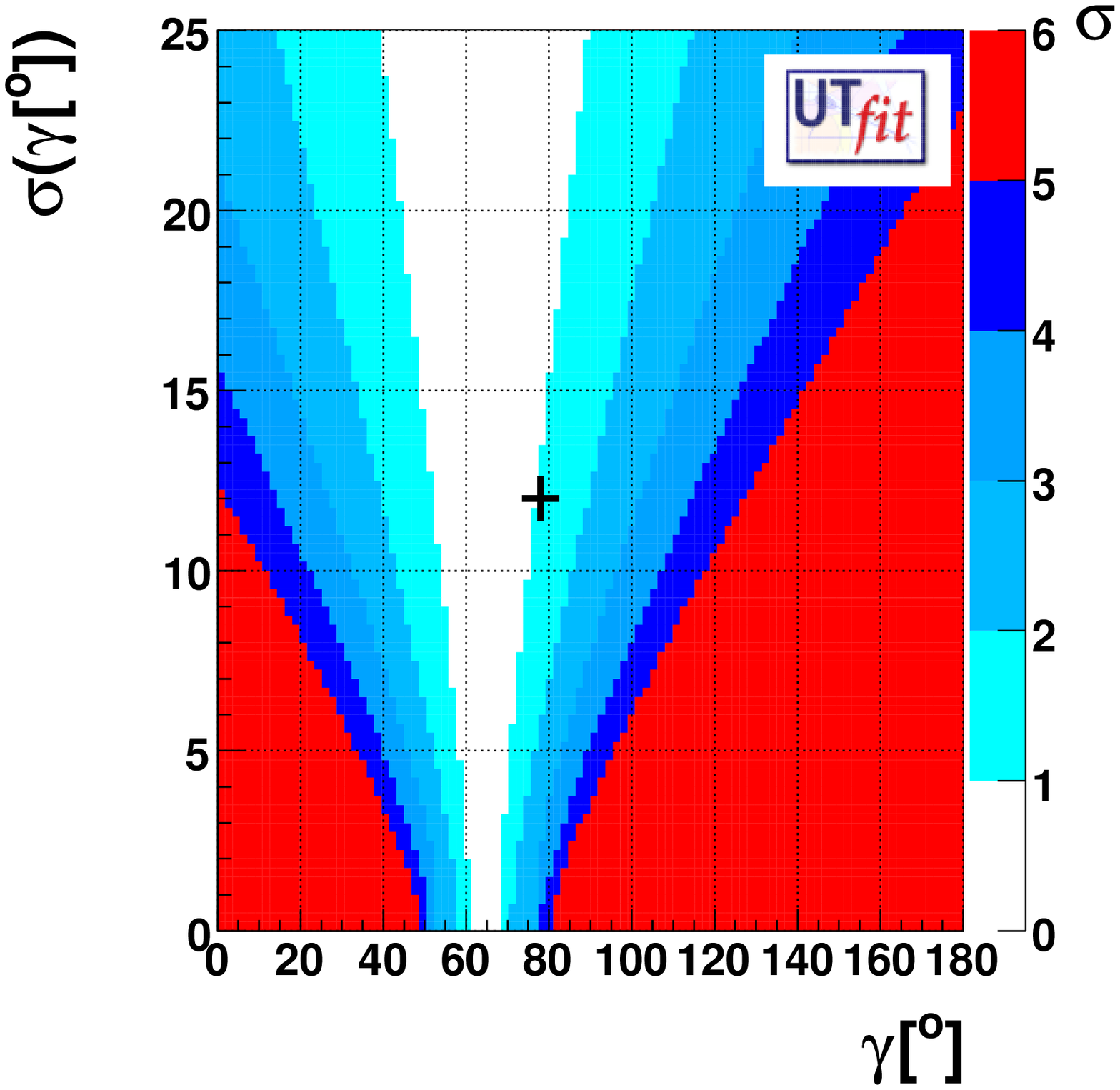,height=3.5cm}
\epsfig{file=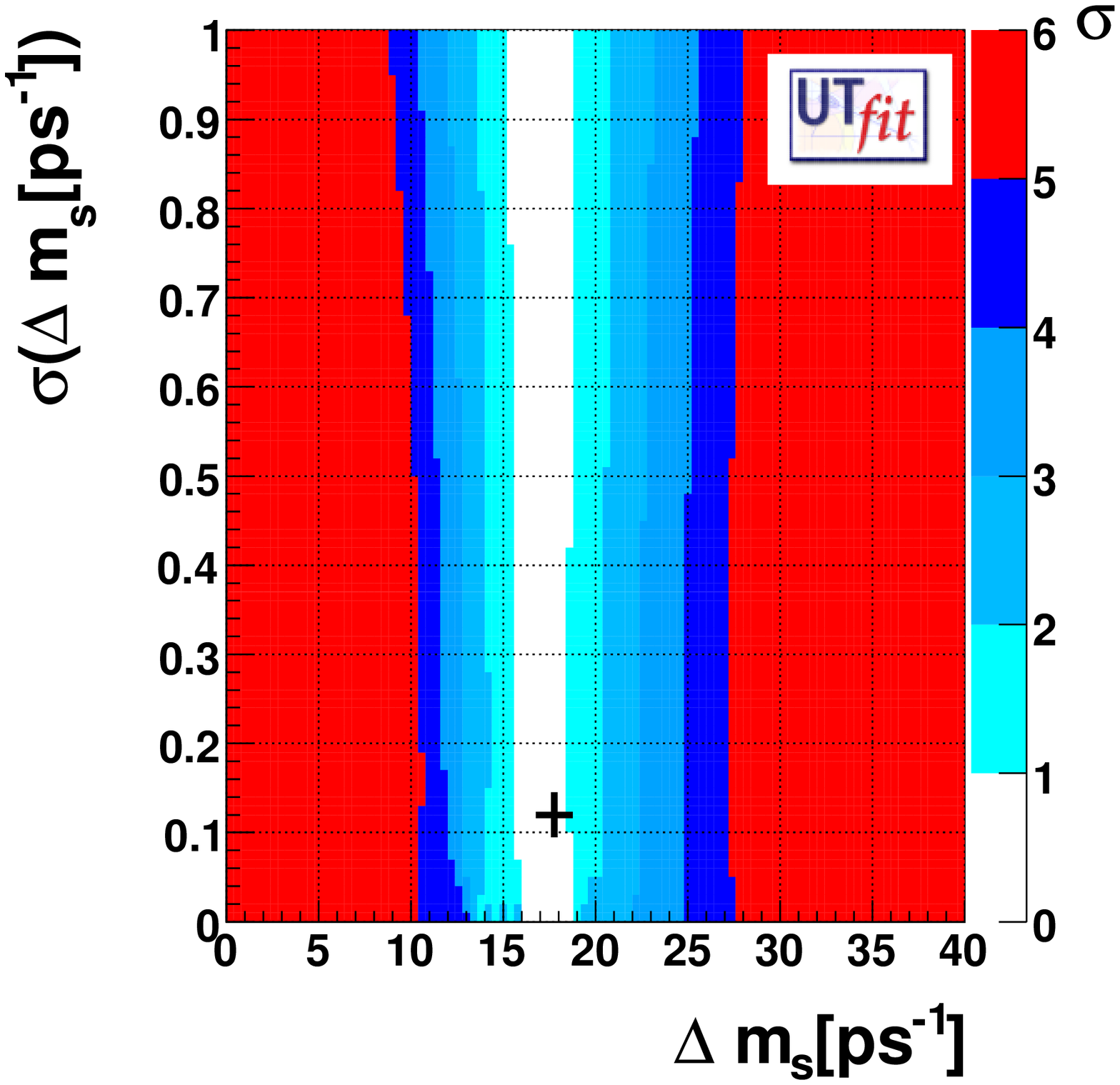,height=3.5cm}
\end{center}
\caption{Compatibility plots for $\alpha$, $\sin 2\beta$, $\gamma$ and $\Delta {m}_s$. 
The color code indicates the compatibility between direct 
and indirect determinations, given in terms of standard deviations, as a function of the 
measured value and its experimental uncertainty. The crosses indicate 
the direct world average measurement values respectively for $\alpha$, $\sin 2\beta$ from 
the measurement of the \CP asymmetry in $B^0 \to J/\psi K^0$, $\gamma$ and 
$\Delta {m}_s$.\label{fig:Ut_various_compatibility}}
\end{figure}

\begin{figure}[h]
\begin{center}
\epsfig{file=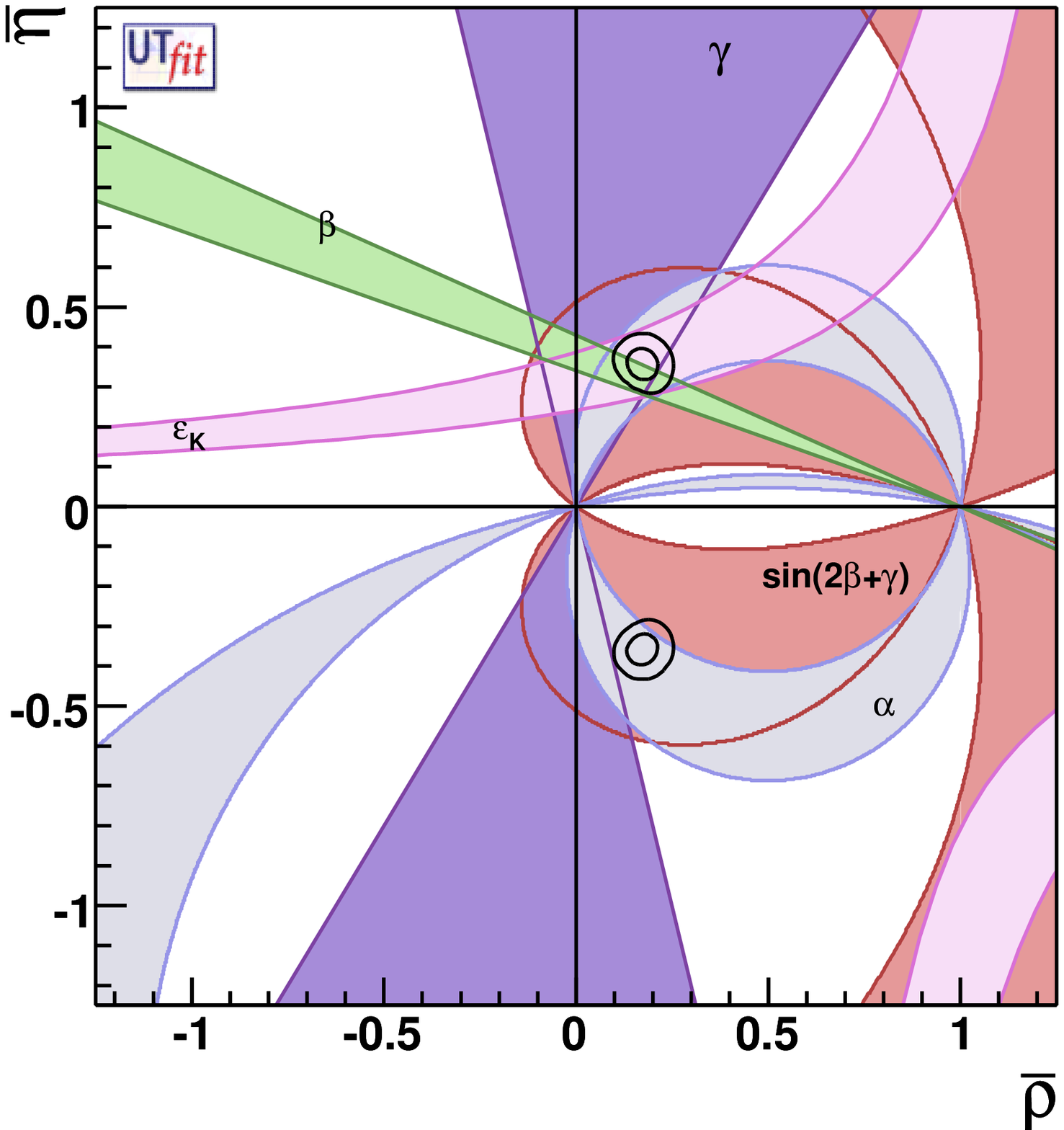,height=5.cm}
\epsfig{file=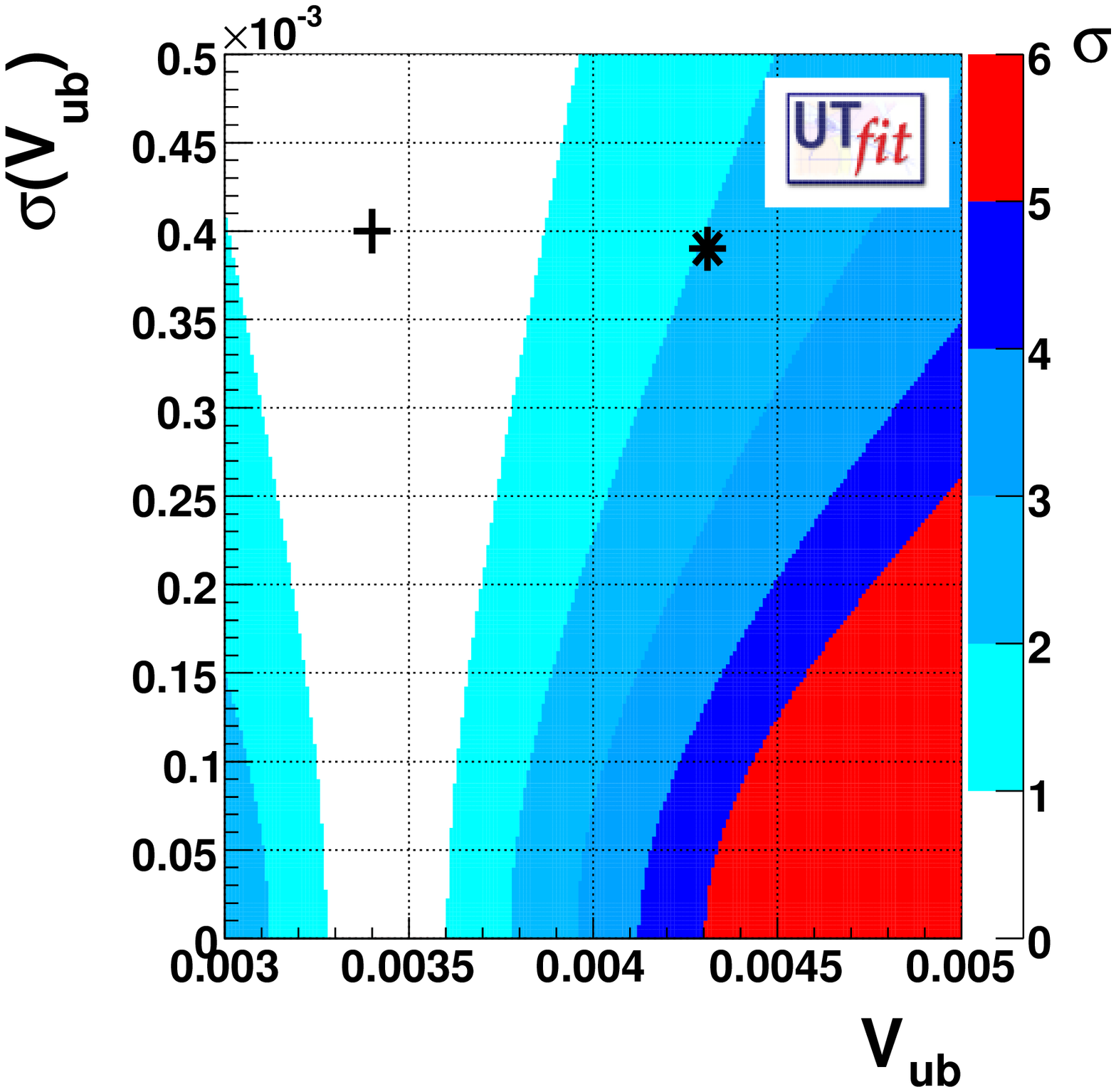,height=5.cm}
\end{center}
\caption{Left plot: allowed regions for $\bar\rho$ and $\bar\eta$ obtained by 
using the measurements of $|\Vub|/|\Vcb|$, $\Delta {m}_d$, $\Delta {m}_s$ and 
$\epsilon_K$.  The coloured zones 
indicate the 68$\%$ and 95$\%$ probability regions for the angles measurements, 
which are not included in the fit.  Right plot: compatibility plot for $V_{ub}$. 
The color code indicates the compatibility between direct and indirect determinations, 
given in terms of standard deviations, as a function of the measured value and 
its experimental uncertainty. 
The cross and the star indicate the exclusive and inclusive measurement values 
respectively\label{fig:Ut_Vub_compatibility}}
\end{figure}

The compatibility plots for $\alpha$, $\sin 2\beta$, $\gamma$ and $\Delta {m}_s$ are shown 
in Fig. \ref{fig:Ut_various_compatibility}.
The direct values obtained for $\alpha$ and $\Delta {m}_s$ are in very good agreement, within 1$\sigma$, 
with the indirect determination, although for the latter the effectiveness of the comparison 
is limited by the precision on the theoretical inputs, inducing a big error (compared to the 
experimental one) on the prediction from the rest of the fit. 
The determination of $\gamma$ from direct measurement yields a value slightly 
higher, $(78\pm 12)^{o}$, than the indirect one from the overall fit, $(64\pm 3)^{o}$; the 
two determinations are compatible within 1$\sigma$.
We also observe that the direct determination of $\sin 2\beta$ from the measurement 
of the \CP asymmetry in $B^0 \to J/\psi K^0$ is slightly shifted, with respect 
to the indirect determination, still being compatible with it within 2$\sigma$.
This effect is visually evident in Fig.~\ref{fig:Ut_Vub_compatibility}, left, where 
the $68\%$ and $95\%$ probability regions for $\bar{\rho}$ and $\bar{\eta}$, as 
given by $|V_{ub}|/|V_{cb}|$, $\Delta m_d$, $\Delta m_s$ and $\epsilon_K$ are compared 
with the $95\%$ probability regions given by the measurements of angles.

This slight tension (see Fig.~~\ref{fig:Ut_Vub_compatibility}, left plot) in the 
UT fit can be related \cite{Bona:2006ah} 
to the fact that the present experimental measurement of 
$\sin 2\beta$ favours a value of $|V_{ub}|$ that is more compatible with the direct 
determination of $|V_{ub}|$ using exclusive methods rather than the one obtained 
using the inclusive ones.  
In Fig.~\ref{fig:Ut_Vub_compatibility}, right, we show the compatibility for the 
exclusive and the inclusive direct determination of $|V_{ub}|$.

\section{Unitarity Triangle analysis and theoretical inputs}

Given the abundance of constraints now available for the determination 
of the CKM parameters, $\bar\rho$ and $\bar\eta$, we can remove from the 
fitting procedure the hadronic parameters coming from lattice. 
In this way we can compare the uncertainty obtained on a given quantity through 
the UT fit to the present theoretical error on the same quantity. The aim of this exercise 
is to quantify the impact that eventual improvements on the lattice calculation 
errors will have on the UT analysis. 

\begin{figure}[h]
\begin{center}
\epsfig{file=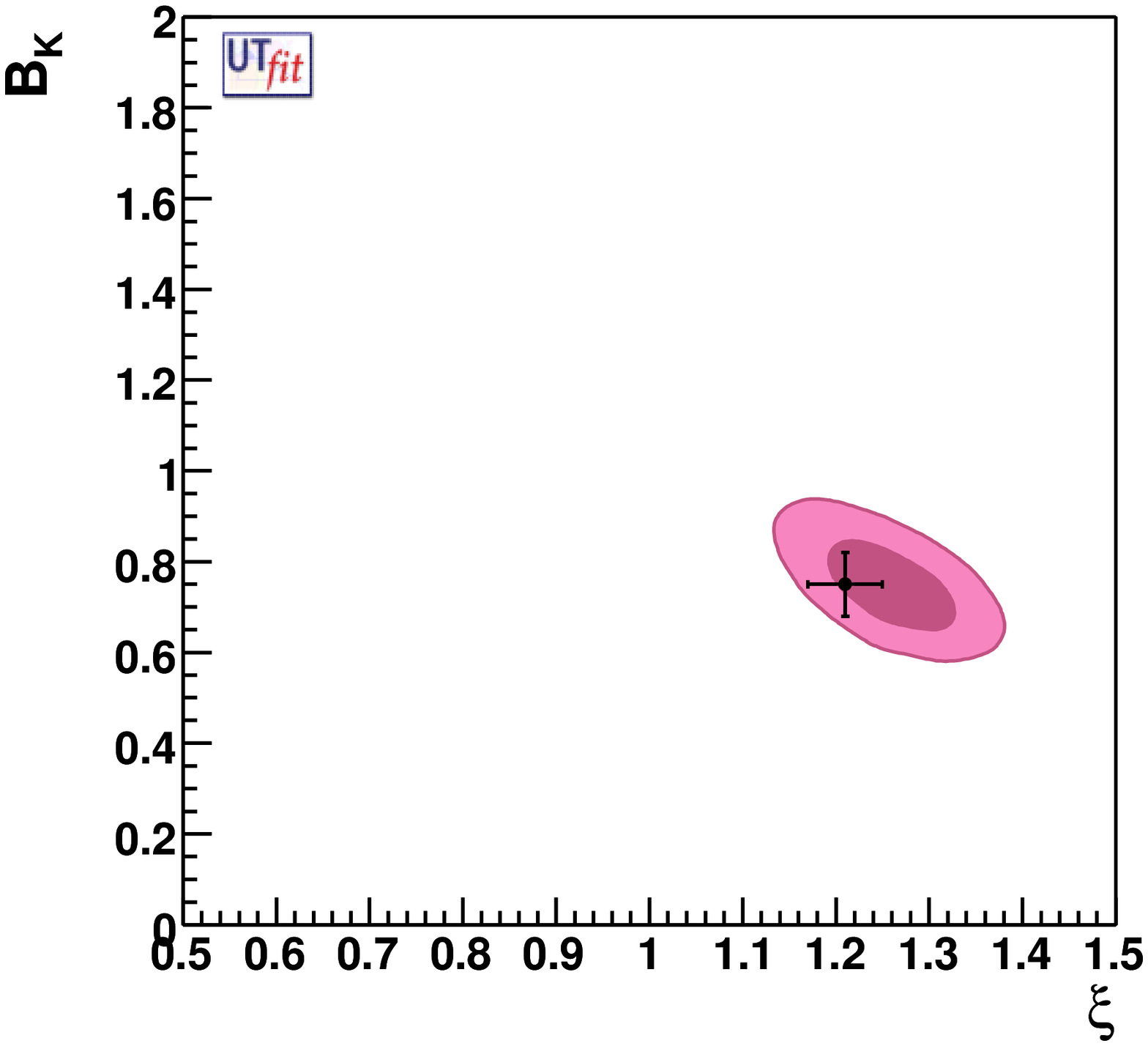,height=5.cm}
\epsfig{file=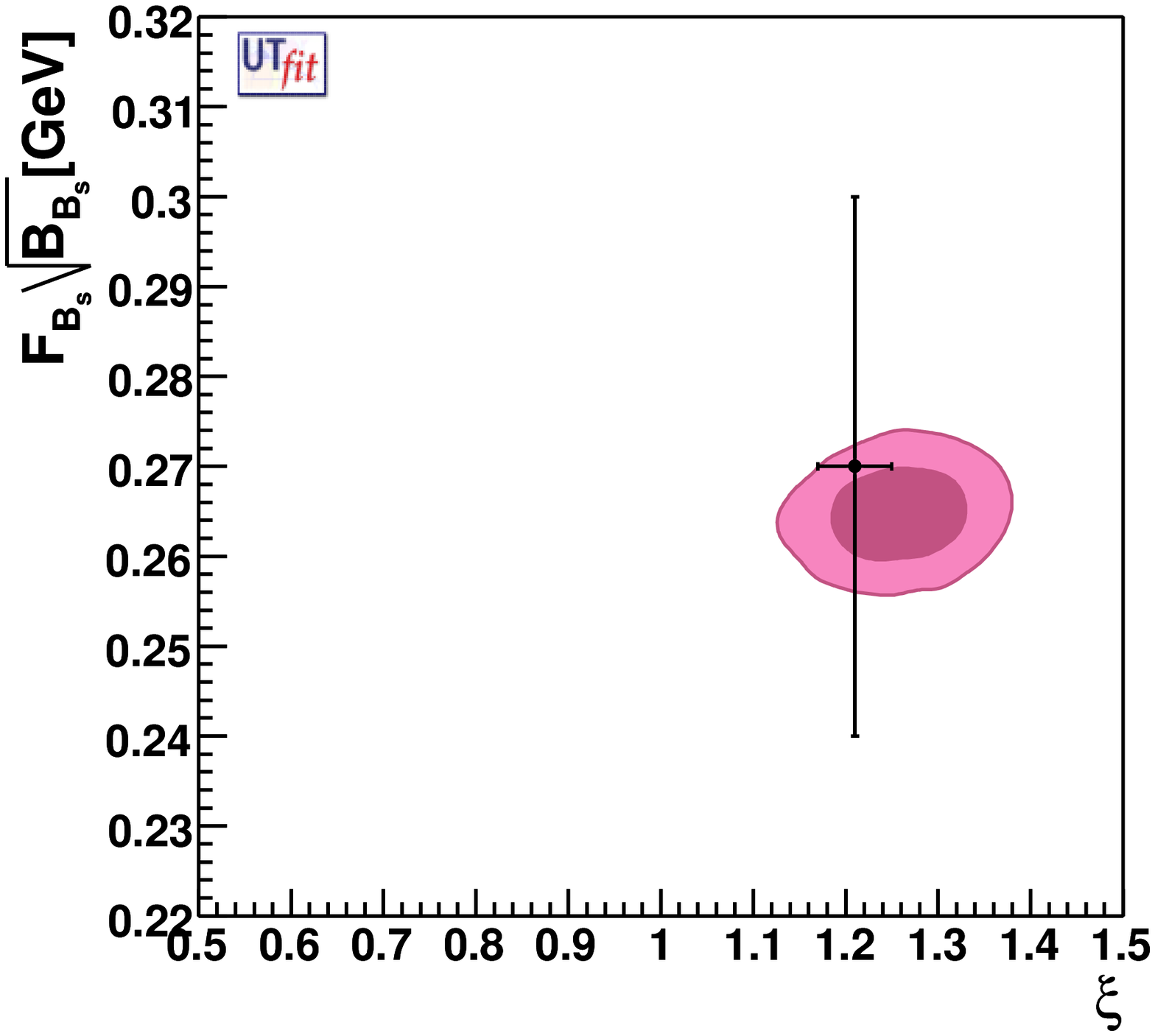,height=5.cm}
\end{center}
\caption{The dark and light colored areas show the 68$\%$ and 95$\%$ probability 
regions in the 2-dimensional plane ($\xi$, $B_{K}$) (left plot) and 
($\xi$, $f_{B_{s}}\sqrt{B_{B_{s}}}$) (right plot). 
The points with error bars show the results of lattice calculations. 
\label{fig:Ut_vs_lattice}}
\end{figure}

In Fig.~\ref{fig:Ut_vs_lattice}, we show the 68$\%$ and 95$\%$ probability 
regions for different lattice quantities, obtained from a UT fit using 
all the measurements of angles and the constraints coming from semileptonic 
$B$ decays. 
The relations between observables and theoretical quantities used in this 
fit are obtained assuming the validity of the SM. 
Numerical results are given in table \ref{tab:Ut_vs_lattice}. 

\begin{table*}[h]
\begin{center}
\begin{tabular}{|l||c|c|}
\hline
Parameter                      & UT (angles+$V_{ub}/V_{cb}$) & Lattice QCD results \\
\hline
$B_{K}$                         & $0.75 \pm 0.07$ & $0.75 \pm 0.07$\\
$f_{B_{s}}\sqrt{B_{B_{s}}}$ [MeV] & $264.7 \pm 3.6$  & $270 \pm 30$   \\
$\xi$                           & $1.26 \pm 0.05$ & $1.21 \pm 0.04$ \\
$f_{B_{d}}$ [MeV]                & $191 \pm 13$     & $200 \pm 20$ \\
\hline
\end{tabular}
\end{center}
\caption{The values obtained for the theoretical parameters from a UT analysis 
using the angles and $V_{ub}/V_{cb}$ measurements are compared with the results 
of lattice calculations.}
\label{tab:Ut_vs_lattice}
\end{table*}


\section{Results of the Unitarity Triangle Analysis beyond the Standard Model}
\label{sec:NP}

Thanks to the abundance of experimental information, the UT analyses can put 
bounds to NP parameters, simultaneously to the determination of the CKM ones. 
We do not consider here NP models with large tree-level effects. 
The starting point for such studies is a New Physics (NP) free determination 
of $\bar\rho$ and $\bar\eta$, in which we only use the constraints from the 
angle $\gamma$ and the semileptonic $B$ decays. These quantities are measured 
from the study of decay channels that proceed only through tree amplitudes and 
can hence be considered free from NP contributions.

The possible contributions of NP effects to $K \bar{K}$, $B_d \bar{B}_d$ and $B_s \bar{B}_s$ 
mixing are then parametrized in a model-independent way in terms of only two 
parameters describing the difference, respectively in absolute value and phase, of 
the amplitude with respect to the SM one. 
In the case of $B_q \bar{B}_q$ mixing ($q=d, s$), the two parameters $C_{{B}_q}$ 
and $\phi_{{B}_q}$ are defined as follows:
\begin{eqnarray*}
C_{{B}_q} e^{2 i \phi_{{B}_q}} = \frac{< {B}_q | H_{eff}^{full}| \bar{B}_q> }{< {B}_q | H_{eff}^{SM}| \bar{B}_q >} = 
\frac{A_{q}^{SM} e^{2i\phi_q^{SM}} + A_{q}^{NP} e^{2i(\phi_q^{NP} + \phi_q^{SM})}    }{ A_{q}^{SM} e^{2i\phi_q^{SM}} }
\end{eqnarray*}

The case of the SM is recovered for $C_{{B}_q}=1$ and $\phi_{{B}_q}=0$ and any 
significant deviation from these values is an indication of the presence of NP. 
The advantage of this parametrization is the factorization of the sources 
of errors, $C_{{B}_q}$ ($\phi_{{B}_q}$) error being determined from the 
theoretical (experimental) precision.  
In the case of $K$-$\bar{K}$ mixing, two parameters can be defined in a similar 
way: 
\begin{eqnarray*}
C_{\Delta m_K} = \frac{Re[<K^0|H_{eff}^{full}|\bar{K}^0>]}{Re[<K^0|H_{eff}^{SM}|\bar{K}^0>]} \; , \; C_{\epsilon_K} = \frac{Im[<K^0|H_{eff}^{full}|\bar{K}^0>]}{Im[<K^0|H_{eff}^{SM}|\bar{K}^0>]} .
\end{eqnarray*}

How the expression of the different observable change in this generic NP scenario 
is described in table \ref{tab:NP_obsChange} (as explained $\gamma$, $V_{ub}$ and $V_{cb}$ 
stay unchanged). 

\begin{table*}[h]
\begin{center}
\begin{tabular}{|c||c|c|c|c|c|c|c|}
\hline
SM & $\epsilon_K^{SM}$ & $\Delta m_K^{SM}$ & $\beta^{SM}$ & $\alpha^{SM}$ & $\Delta m_d^{SM}$ & $\beta_s^{SM}$ & $\Delta m_s^{SM}$\\\hline\hline
SM + NP & $C_{\epsilon_K} \epsilon_K^{SM}$ & $C_{\Delta m_K} \Delta m_K^{SM}$ & $\beta^{SM}+\phi_{B_d}$ & $\alpha^{SM}-\phi_{B_d}$ & $C_{B_d}\Delta m_d^{SM}$ & $\beta_s^{SM}-\phi_{B_s}$ & $C_{B_s}\Delta m_s^{SM}$\\
\hline
\end{tabular}
\end{center}
\caption{Expression for the different observables in the model-independent 
NP parametrization.}
\label{tab:NP_obsChange}
\end{table*} 

The additional experimental inputs used for this analysis are listed in 
table \ref{tab:exp_inputsNP}.

\begin{table*}[h]
\begin{center}
\begin{tabular}{|l||c|c|c|}
\hline
Input & Source & Value & Reference \\
\hline
$A_{SL}$ [$\times 10^{2}$]&   Semileptonic $B_s$ decays     & $-0.20 \pm 1.19$ & \cite{ref:D0Asl_s} \\
$A_{\mu\mu}$ [$\times 10^{3}$]  & $p\bar{p} \to \mu\mu X$  & $-4.3 \pm 3.0$  &\cite{ref:AmumuD0,ref:AmumuCDF} \\
$\tau^{FS}_{B_s}$ [ps] & Flavor specific $B_s$ final states       & $1.461 \pm 0.032$ & \cite{ref:tauBsFS} \\
$\Delta\Gamma_s$,$\phi_s$ & $B_s\to J/\Psi \phi$  & 2-dimensional likelihoods  & \cite{ref:DGsPhis_D0,ref:DGsPhis_CDF} \\
\hline
\end{tabular}
\end{center}
\caption{Additional inputs to the UT fit for NP analyses.}
\label{tab:exp_inputsNP}
\end{table*}

\begin{figure}[h]
\begin{center}
\epsfig{file=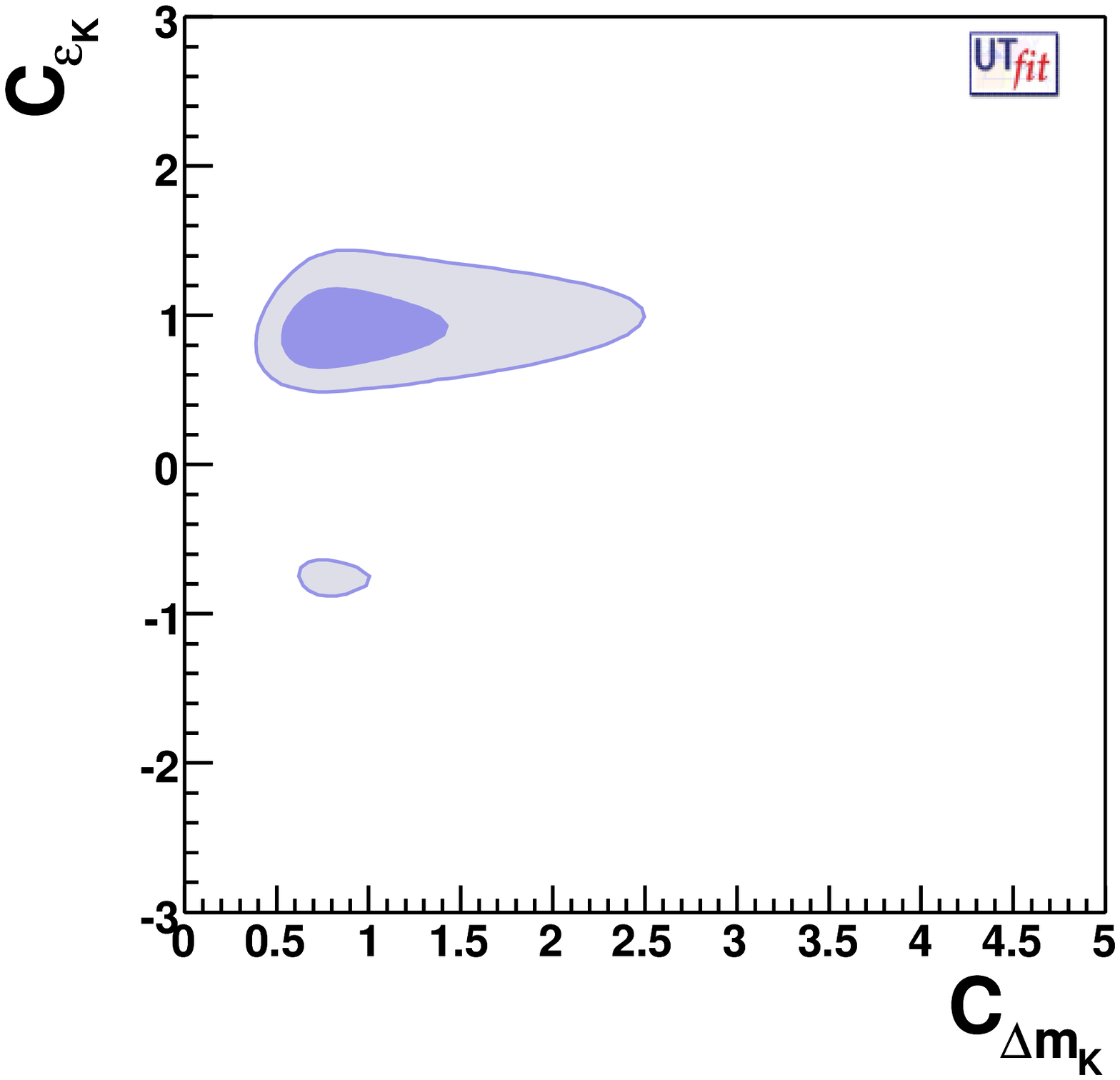,height=5.cm}
\epsfig{file=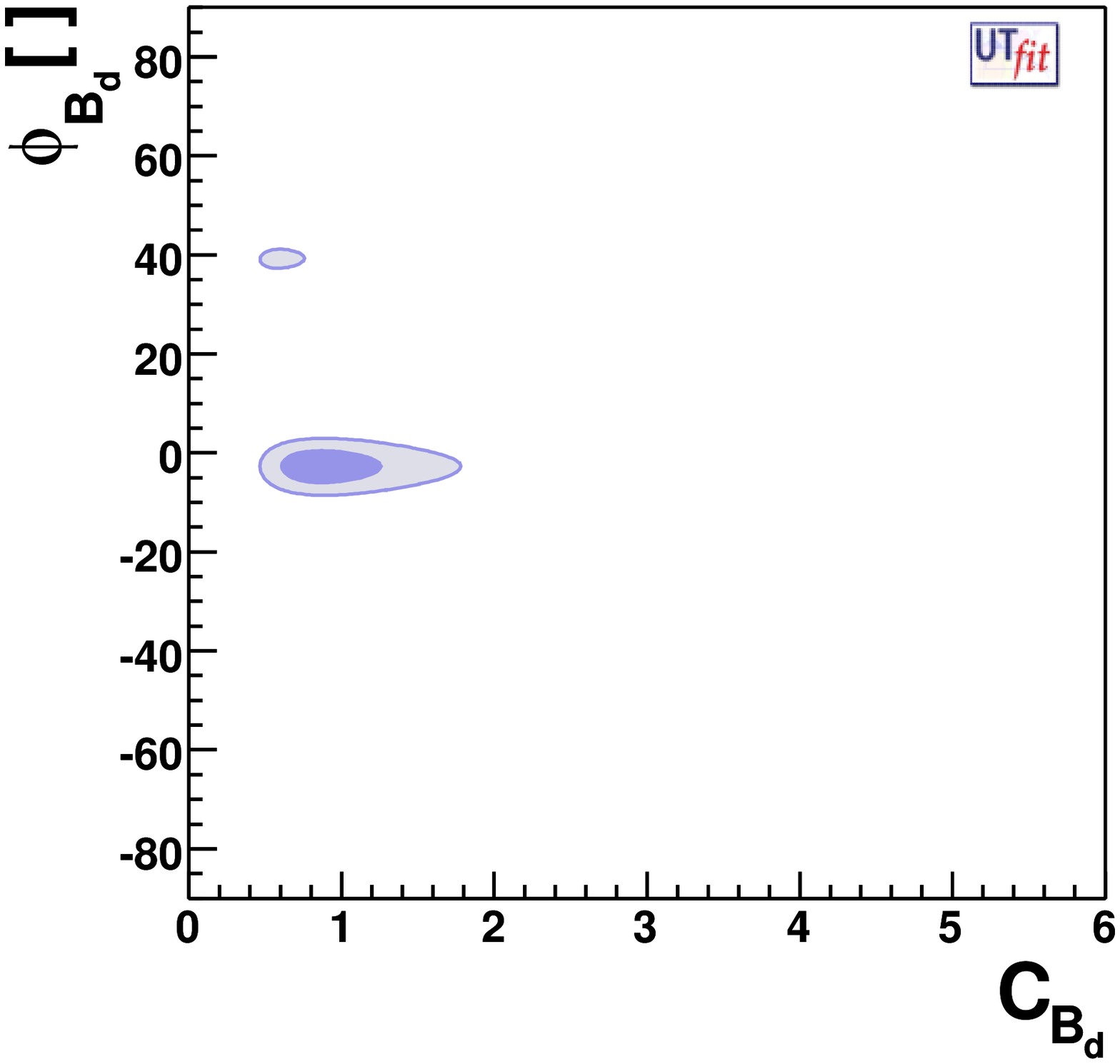,height=5.cm}
\epsfig{file=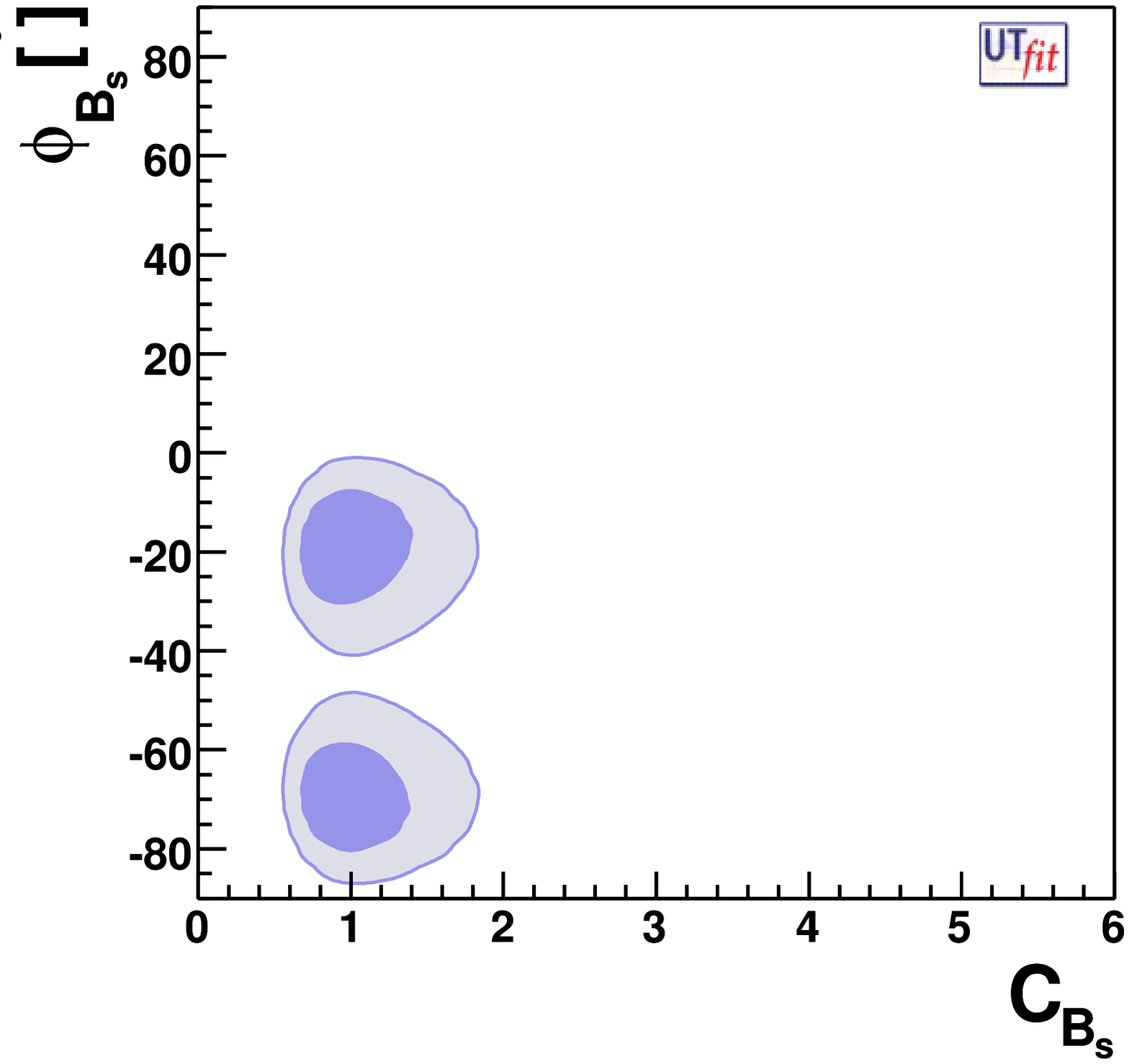,height=5.cm}
\end{center}
\caption{The dark and light colored areas show the 68$\%$ and 95$\%$ probability 
regions in the 2-dimensional plane ($C_{\Delta m_K}$, $C_{\epsilon_K}$) (left plot), 
($C_{B_d}$,$\phi_{B_d}$) (center plot), ($C_{B_s}$,$\phi_{B_s}$) (right plot).  \label{fig:NP_results}}
\end{figure}

Figure \ref{fig:NP_results} shows the result of the NP analysis for the 
Kaon, $B_d$ and $B_s$ sectors. Numerical results for the additional NP 
parameters are summarized in table \ref{tab:NP_results}. 

\begin{table*}[h]
\begin{center}
\begin{tabular}{|l||c|c|c|c|c|c|}
\hline
Parameter & $C_{\epsilon_K}$ & $C_{\Delta_{m_K}}$ & $C_{B_d}$ & $\phi_{B_d}$ $[^{\circ}]$& $C_{B_s}$ & $\phi_{B_s}$ $[^{\circ}]$\\\hline
Value  & $0.91 \pm 0.13$ & $0.96 \pm 0.34$ & $0.90 \pm 0.23$ & $-2.7 \pm 1.9$ & $0.99 \pm 0.23$ & $(-70\pm 7)$U$(-18\pm 7)$\\
\hline
\end{tabular}
\end{center}
\caption{Numerical results (at 68\% probability) for the NP parameters.}
\label{tab:NP_results}
\end{table*}

Given the experimental measurements, the results for $\phi_{B_s}$ show a discrepancy 
of 2.9$\sigma$ from the SM value, pointing to NP contributions with new sources 
of flavor violation in the transition within 2$^{nd}$ and 3$^{rd}$ generation. 
The results for $\phi_{B_d}$ show a slight discrepancy from the SM value, of the 
order of 1.5 $\sigma$, which is a consequence of the tension in the UT global fit 
mentioned in section \ref{sec:compatibility}. As a consequence, NP contributions in 
transitions within 1$^{st}$ and 3$^{rd}$ generations are 
not yet excluded, but are limited to be of the order of 10\% at most. 
Finally, generic NP contributions in the transitions within 1$^{st}$ and 2$^{nd}$ 
generations are strongly suppressed \footnote{The corrections to $\epsilon_K$, 
discussed in \cite{ref:Guadagnoli}, are not yet included in this analysis and 
could give sizeable effects}.




\begin{thebibliography}{99}
\bibitem{ref_dms} For this only the CDF measurement (Phys. Rev. Lett.
{\bf 97}, 242003 (2006)) has been used as it current dominates the 
world average which still needs to be determined by HFAG.
\end{thebibliography}
\end{document}